\documentclass[aps,onecolumn,showpacs,nofootinbib,showkeys]{revtex4-2}
\usepackage[margin=3.15cm]{geometry}

\RequirePackage[T1]{fontenc}

\usepackage{epsfig,graphicx,amsmath,amssymb,bm}
\RequirePackage{mathptmx}      
\usepackage{mathrsfs}
\usepackage{amssymb}
\RequirePackage{color}
\usepackage{changes}
\usepackage{slashed}
\usepackage{multirow}
\usepackage{scalerel}
\usepackage{tikz-feynman}
\usepackage{textcomp}
\usepackage{subcaption}
\usepackage[english]{babel}
\usepackage{float}
\RequirePackage{hyperref}
\hypersetup{
    linktocpage,
    colorlinks,
    citecolor=blue,
    filecolor=black,
    linkcolor=blue,
    urlcolor=blue,
}
\usepackage{nccmath}

\begin{document}

\title{The decays $\tau \to a_{1} \pi \nu_\tau$, $\tau \to K_{1} \pi \nu_\tau$ and $\tau \to K_{1} K \nu_\tau$ in the extended $U(3)\times U(3)$ chiral NJL model}


\author{Mikhail K. Volkov$^{1}$}\email{volkov@theor.jinr.ru}
\author{Aleksey A. Pivovarov$^{1}$}\email{tex$\_$k@mail.ru}
\author{Kanat Nurlan$^{1,2,3}$}\email{nurlan@theor.jinr.ru}

\affiliation{$^1$ Bogoliubov Laboratory of Theoretical Physics, JINR, 
                 141980 Dubna, Moscow region, Russia \\
                $^2$ The Institute of Nuclear Physics, Almaty, 050032, Kazakhstan\\
                $^3$ Al-Farabi Kazakh National University, Almaty, 050040 Kazakhstan}   


\begin{abstract}
The branching fractions of the decays $\tau \to a_{1} \pi \nu_\tau$, $\tau \to K_{1} \pi \nu_\tau$ and $\tau \to K_{1} K \nu_\tau$ are calculated in the framework of the extended $U(3)\times U(3)$ chiral Nambu--Jona-Lasinio model. There are no experimental data for these decays in the present time, thus the obtained results are considered as predictions. The comparison of these results to the theoretical results of other authors is carried out.


\end{abstract}


\maketitle


\section{\label{Intro}Introduction}
Study of the $\tau$ lepton decays into hadrons plays an important role for deeper understanding nature of the strong meson interactions at low energies. This also allows to specify the relation of these interactions to the fundamental QCD theory. By present time, numerous experimental studies of measurement of $\tau$ decays into hadrons have been carried out \cite{ParticleDataGroup:2020ssz}. There are many theoretical works explaining nature of these interactions (\cite{Davier:2005xq, Volkov:2017arr, Volkov:2022jfr} and references in them). The $\tau$ lepton decays with production of two pseudoscalar mesons or one vector and one pseudoscalar meson have been studied quite well \cite{Volkov:2022jfr}. However, the decays of $\tau$ lepton into axial vector and pseudoscalar mesons are significantly less known. There are experimental results only for the process $\tau \to f_1(1285) \pi \nu_\tau$ \cite{ParticleDataGroup:2020ssz}. There are also theoretical studies of this decay \cite{Li:1996md, Calderon:2012zi, Volkov:2018fyv, Oset:2018zgc, Dai:2018zki}. However, the decays $\tau \to a_1(1260) \pi \nu_\tau$, $\tau \to K_1(1270) \pi \nu_\tau$, $\tau \to K_1(1400) \pi \nu_\tau$, $\tau \to K_1(1270) K \nu_\tau$ are still less known. By present time, there are no experimental data for these processes, but there are some articles describing these processes in various theoretical approaches. For example, in the work \cite{Calderon:2012zi}, the calculation have been carried out in the model based on the vector dominance. In the work \cite{Dai:2018zki}, the unitary extensions of chiral perturbation theory (U$\chi$PT) has been applied where axial vector mesons are produced in the vector-pseudoscalar scattering amplitudes. However, the results obtained in these works significantly differ from each other. That is why, it seems to be interesting to carry out the appropriate calculations in the framework of the $U(3) \times U(3)$ chiral Nambu--Jona-Lasinio (NJL) model \cite{Volkov:2022jfr, Volkov:2005kw, Vogl:1991qt, Ebert:1994mf} which differs from the models applied in the previous works. This problem is considered in the present paper.   

\section{Quark-meson Lagrangian of the extended NJL model}
    In the extended NJL model, the Lagrangian describing interactions of the considered pseudoscalar, vector and axial vector mesons with quarks takes the form \cite{Volkov:2022jfr, Volkov:2005kw}:
    \begin{eqnarray}
    \label{Lagrangian}
    	\Delta L_{int} & = & \bar{q}\sum_{i=0,\pm}\left[iA_{\pi}\gamma^{5}\lambda^{\pi}_{i}\pi^{i} +
    	iA_{K}\gamma^{5}\lambda^{K}_{i}K^{i} + \frac{1}{2}\gamma^{\mu}\lambda^{\rho}_{i}(A_{\rho}\rho^{i}_{\mu} + B_{\rho}\rho^{'i}_{\mu}) \right. \nonumber\\
    	&&\left. + \frac{1}{2}\gamma^{\mu}\lambda^{K}_{i}(A_{K^{*}}K^{*i}_{\mu} + B_{K^{*}}K^{*'i}_{\mu}) + \frac{A_{K_{1}}}{2}\gamma^{\mu}\lambda^{K}_{i}K^{i}_{1\mu} + \frac{A_{a_{1}}}{2}\gamma^{\mu}\lambda^{\rho}_{i}a^{i}_{1\mu}\right]q,
	\end{eqnarray}
	where $q$ and $\bar{q}$ are the fields of u-, d- and s-quarks with the constituent masses $m_{u} = m_{d} = 270$~MeV, $m_{s} = 420$~MeV; the matrices $\lambda$ are the linear combinations of the Gell-Mann matrices. The first radially excited meson states are marked with prime. The factors $A$ and $B$, appearing as a result of diagonalization of the initial Lagrangian take the form \cite{Volkov:2022jfr}
	\begin{eqnarray}
	\label{verteces1}
    	A_{M} & = & \frac{1}{\sin(2\theta_{M}^{0})}\left[g_{M}\sin(\theta_{M} + \theta_{M}^{0}) +
    	g_{M}^{'}f_{M}(k_{\perp}^{2})\sin(\theta_{M} - \theta_{M}^{0})\right], \nonumber\\
    	B_{M} & = & \frac{-1}{\sin(2\theta_{M}^{0})}\left[g_{M}\cos(\theta_{M} + \theta_{M}^{0}) +
    	g_{M}^{'}f_{M}(k_{\perp}^{2})\cos(\theta_{M} - \theta_{M}^{0})\right].
    \end{eqnarray}
    The index $M$ specifies a corresponding meson.
        The mixing angles $\theta_{M}$ appearing as a result of the diagonalization of the initial Lagrangian:
    \begin{eqnarray}
    \label{angels}
    	&\theta_{\rho} = 81.8^{\circ}, \quad \theta_{K} = 58.11^{\circ}, \quad \theta_{K^{*}} = 84.74^{\circ}, \quad \theta_{a_{1}} = \theta_{\rho}, \quad \theta_{K_{1}} = \theta_{K^{*}},& \nonumber\\
    	&\theta_{\rho}^{0} = 61.5^{\circ}, \quad \theta_{K}^{0} = 55.52^{\circ}, \quad \theta_{K^{*}}^{0} = 59.56^{\circ}, \quad \theta_{a_{1}}^{0} = \theta_{\rho}^{0}, \quad \theta_{K_{1}}^{0} = \theta_{K^{*}}^{0}.&
    \end{eqnarray}
    In the case of the pions one can assume $\theta_{\pi} \approx \theta_{\pi}^{0} \approx 59.12^{\circ}$. 
    
    The form factor applied to include the first radially excited meson states takes the form
    \begin{eqnarray}
        f\left(k_{\perp}^{2}\right) = \left(1 + d k_{\perp}^{2}\right)\Theta(\Lambda^{2} - k_{\perp}^2),
    \end{eqnarray}
    where $d$ is the slope parameter depending only of the quark composition of a meson \cite{Volkov:2022jfr}, $k$ is relative momentum of quarks in a meson.
    
       The coupling constants:
    \begin{eqnarray}
    \label{Couplings}
    	g_{\rho} = g_{a_{1}} =  \left(\frac{3}{2I_{20}}\right)^{1/2}, &\quad& g_{\rho}^{'} = g_{a_{1}}^{'} =  \left(\frac{3}{2I_{20}^{f^{2}}}\right)^{1/2}, \nonumber\\
    	g_{K^{*}} = g_{K_{1}} = \left(\frac{3}{2I_{11}}\right)^{1/2}, &\quad& g_{K^{*}}^{'} = g_{K_{1}}^{'} =  \left(\frac{3}{2I_{11}^{f^{2}}}\right)^{1/2}, \nonumber\\
    	g_{K} =  \left(\frac{Z_{K}}{4I_{11}}\right)^{1/2}, &\quad& g_{K}^{'} =  \left(\frac{1}{4I_{11}^{f^{2}}}\right)^{1/2}, \nonumber\\
    	g_{\pi} =  \left(\frac{Z_{\pi}}{4I_{20}}\right)^{1/2}, &\quad& g_{\pi}^{'} =  \left(\frac{1}{4I_{20}^{f^{2}}}\right)^{1/2},
    \end{eqnarray}
    where $Z_{\pi}$ and $Z_{K}$ are the additional renormalization constants appearing when taking into account $\pi-a_{1}$ and $K-K_{1}$ transitions:
    \begin{eqnarray}
        Z_{\pi} & = & \left(1 - 6 \frac{m_{u}^{2}}{M_{a_{1}}^{2}}\right)^{-1}, \nonumber\\
    	Z_{K} & = & \left(1 - \frac{3}{2} \frac{(m_{u} + m_{s})^{2}}{M_{K_{1A}}^{2}}\right)^{-1}, \nonumber\\
    	M_{K_{1A}} & = & \left(\frac{\sin^{2}{\alpha}}{M^{2}_{K_{1}(1270)}} + \frac{\cos^{2}{\alpha}}{M^{2}_{K_{1}(1400)}}\right)^{-1/2}.
    \end{eqnarray}
    The split of the state $K_{1A}$ into two physical mesons $K_{1}(1270)$ and $K_{1}(1400)$ with the mixing angle $\alpha \approx 57^{\circ}$  \cite{Suzuki:1993yc, Volkov:2019awd} is taken into account here.
    The masses of these states are $M_{K_{1}(1270)} = 1253 \pm 7$~MeV and $M_{K_{1}(1400)} = 1403 \pm 7$~MeV~\cite{ParticleDataGroup:2020ssz}. 
    
    The integrals appearing in definition of the coupling constants appear in the quark loops as a result of the renormalization of the Lagrangian are:
    \begin{eqnarray}
    	I_{n_{1}n_{2}}^{f^{m}} =
    	-i\frac{N_{c}}{(2\pi)^{4}}\int\frac{f^{m}(k_{\perp}^{2})}{(m_{u}^{2} - k^2)^{n_{1}}(m_{s}^{2} - k^2)^{n_{2}}}\Theta(\Lambda^2 - k_{\perp}^{2})
    	\mathrm{d}^{4}k,
    \end{eqnarray}
    where $\Lambda = 1.03$~GeV is a cut-off parameter~\cite{Volkov:2022jfr}. 

\section{Amplitudes and numerical estimates} 
\begin{figure*}[t]
 \centering
  \begin{subfigure}{0.5\textwidth}
   \centering
    \begin{tikzpicture}
     \begin{feynman}
      \vertex (a) {\(\tau\)};
      \vertex [dot, right=2cm of a] (b){};
      \vertex [above right=2cm of b] (c) {\(\nu_{\tau}\)};
      \vertex [dot, below right=1.2cm of b] (d) {};
      \vertex [dot, above right=1.2cm of d] (e) {};
      \vertex [dot, below right=1.2cm of d] (h) {};
      \vertex [right=1.2cm of e] (f) {\(a_1\)};
      \vertex [right=1.2cm of h] (k) {\(\pi\)}; 
      \diagram* {
         (a) -- [fermion] (b),
         (b) -- [fermion] (c),
         (b) -- [boson, edge label'=\(W\)] (d),
         (d) -- [fermion] (e),  
         (e) -- [fermion] (h),
         (h) -- [anti fermion] (d),
         (e) -- [] (f),
         (h) -- [] (k),
      };
     \end{feynman}
    \end{tikzpicture}
   \caption{$\tau \to a_1 \pi \nu_\tau$ decay contact diagram}
  \end{subfigure}%
 \centering
 \begin{subfigure}{0.5\textwidth}
  \centering
   \begin{tikzpicture}
    \begin{feynman}
      \vertex (a) {\(\tau\)};
      \vertex [dot, right=2cm of a] (b){};
      \vertex [above right=2cm of b] (c) {\(\nu_{\tau}\)};
      \vertex [dot, below right=1.2cm of b] (d) {};
      \vertex [dot, right=1cm of d] (l) {};
      \vertex [dot, right=1.2cm of l] (g) {};
      \vertex [dot, above right=1.2cm of g] (e) {};
      \vertex [dot, below right=1.2cm of g] (h) {};
      \vertex [right=1.2cm of e] (f) {\(a_1\)};
      \vertex [right=1.2cm of h] (k) {\(\pi\)}; 
      \diagram* {
         (a) -- [fermion] (b),
         (b) -- [fermion] (c),
         (b) -- [boson, edge label'=\(W\)] (d),
         (d) -- [fermion, inner sep=1pt, half left] (l),
         (l) -- [fermion, inner sep=1pt, half left] (d),
         (l) -- [edge label'=\({\rho, \rho'} \)] (g),
         (g) -- [fermion] (e),  
         (e) -- [fermion] (h),
         (h) -- [anti fermion] (g),
         (e) -- [] (f),
         (h) -- [] (k),
      };
     \end{feynman}
    \end{tikzpicture}
   \caption{Diagram with the intermediate vector mesons $\rho$ and $\rho'$ of the decay $\tau \to a_1 \pi \nu_\tau$}
  \end{subfigure}%
 \caption{Diagrams contributing to the decay $\tau \to a_1 \pi \nu_\tau$.}
 \label{diagrams}
\end{figure*}
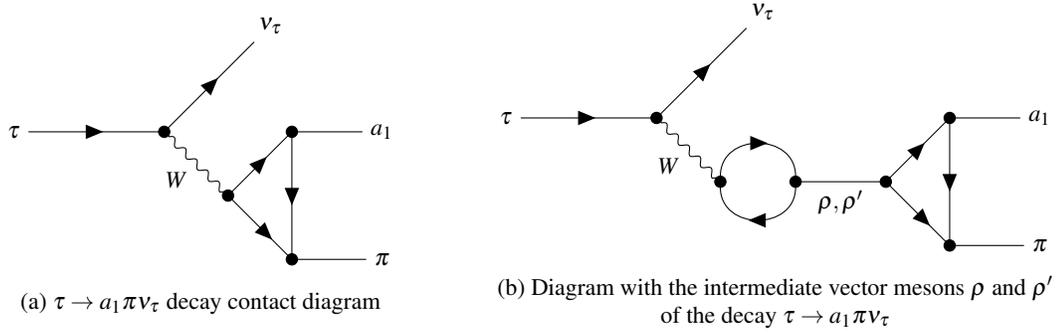

    The decay amplitude of the process $\tau \to a_1(1260) \pi \nu_\tau$ is described by two types of diagrams which are shown in Fig. \ref{diagrams}. The first diagram describes the contact term contribution to the amplitude when W boson decays directly to the final products without any intermediate meson states. The second diagram corresponds to the contributions from channels with intermediate $\rho$ and $\rho'$ vector mesons. 
    
    Note that the quark loop integrals are expanded in momenta of external fields, and only the logarithmic divergent parts are kept. This corresponds to the requirement of conservation of chiral symmetry in the NJL model.
    
    As a result, the amplitude of $\tau \to a_1(1260) \pi \nu_\tau$ decay takes the form
    \begin{eqnarray}
    \label{amplitude_a1pi}
    \mathcal{M}(\tau \to a_1 \pi \nu_\tau) & = & -i G_F V_{ud} 4m_u g_\pi L_\mu \left[ I_{20}^{a_1}g^{\mu\nu} + I_{20}^{a_1 \rho} \frac{C_\rho}{g_\rho} \frac{g^{\mu\nu}s - p^\mu p^\nu}{M^2_\rho - s - i \sqrt{s}\Gamma_\rho} \right. \nonumber\\
    && \left. + \, I_{20}^{a_1 \rho'} \frac{C_{\rho'}}{g_\rho} \frac{g^{\mu\nu}s - p^\mu p^\nu}{M^2_{\rho'} - s - i \sqrt{s}\Gamma_{\rho'}} \right] \epsilon_\nu(p_{a_1}).
    \end{eqnarray} 
    
    Here $G_F = 1.1663787(6)\times 10^{-11}$ $MeV^{-2}$ is the Fermi coupling constant; $V_{ud} = 0.97417 \pm 0.00021$ is the Cabibbo-Kobayashi-Maskawa matrix element; $l_\mu = \bar{\nu_\tau}\gamma_\mu(1-\gamma_5)\tau$ is a lepton current; $\epsilon_\nu (p_{a_1})$ is a polarization vector of the $a_1$ meson with the momentum $p_{a_1}$; $s = {(p_{a_1} + p_{\pi})}^2$ is a square of the invariant mass of the $a_1 \pi$ meson pair, the masses and widths of the mesons are taken form PDG \cite{ParticleDataGroup:2020ssz}. The decays $\tau \to a_1^{-} \pi^{0} \nu_\tau$ and $\tau \to a_1^{0} \pi^{-} \nu_\tau$ are differ in the value of the neutral and charged meson masses.
    
    The constants $C_V$ and $C_{V'}$ describe the $W \to \rho (\rho')$ transitions through the quark loop
         \begin{eqnarray}
        \label{C_const}
            C_{M}= \frac{1}{\sin{\left(2\theta_{M}^{0}\right)}} \left[\sin{\left(\theta_{M} + \theta_{M}^{0}\right)} + R_{M} \sin{\left(\theta_{M} - \theta_{M}^{0}\right)}\right], \\ \nonumber
        C_{M'}= \frac{-1}{\sin{\left(2\theta_{M}^{0}\right)}} \left[\cos{\left(\theta_{M} + \theta_{M}^{0}\right)} + R_{M} \cos{\left(\theta_{M} - \theta_{M}^{0}\right)}\right],    
\end{eqnarray}
where $M$ is the vector meson ($\rho$ meson im this case), $R_{\rho}$ can be defined in the following way
\begin{eqnarray}
    R_{\rho} = \frac{I_{20}^{f}}{\sqrt{I_{20}I_{20}^{ff}}}.
\end{eqnarray}

The integrals
\begin{eqnarray}
\label{integral}
&& I_{n_1n_2}^{M... M'...}(m_{u}, m_{s}) = -i\frac{N_{c}}{(2\pi)^{4}} 
 \int\frac{A(k_{\perp}^{2})...B(k_{\perp}^{2})...}{(m_{u}^{2} - k^2)^{n_1}(m_{s}^{2} - k^2)^{n_2}}
\theta(\Lambda^2 - k_{\perp}^{2}) \mathrm{d}^{4}k,
\end{eqnarray}
are obtained from the quark triangular loops; $A(k_{\perp}^{2})$ and $B(k_{\perp}^{2})$ are the coefficients for different mesons defined in (\ref{verteces1}).

The decay width of $\tau \to a_1 \pi \nu_\tau$ can be calculated by the following formula:
\begin{eqnarray}
\Gamma (\tau \to a_1 \pi \nu_\tau) = \frac{1}{2J_{\tau}+1} \cdot \frac{1}{256 {\pi}^3 M^3_\tau} \int\limits_{s_{-}}^{s_{+}} ds \int\limits_{t_{-}(s)}^{t_{+}(s)} dt \: {|\mathcal{M}(\tau \to a_1 \pi \nu_\tau)|}^2,
	\end{eqnarray}
where the limits of integration have the form
	\begin{eqnarray}
s_+ = M^2_\tau, \quad s_- = (M_{a_1}+M_\pi)^2,
	\end{eqnarray}
 	\begin{eqnarray}
t_{\pm}(s) = \frac{1}{2} \left( M^2_\tau + M^2_{a_1} + M^2_\pi - s + \frac{M^2_\tau}{s} (M^2_{a_1}-M^2_\pi) \pm \sqrt{\Omega(s)}\right),
	\end{eqnarray}
	where
 \begin{eqnarray}
\Omega(s) = s^{-2} \left(s-M^2_\tau \right) \cdot \left(s-(M_{a_1}+M_\pi)^2\right) \cdot \left[ s^2 - s \left(M^2_\tau+(M_{a_1}-M_\pi)^2 \right) + M^2_\tau (M_{a_1}-M_\pi)^2 \right]. \nonumber
	\end{eqnarray}

Let us now consider the $\tau$ decays with the production of strange axial vector mesons $\tau \to K_1(1270) \pi \nu_\tau$, $\tau \to K_1(1400) \pi \nu_\tau$ and $\tau \to K_1(1270) K \nu_\tau$. An important feature of the calculations of these decays is the splitting of the axial vector meson into two physical states $K_1(1270)$ and $K_1(1400)$. This is a consequence of the mixing of the axial-vector states $K_{1A}$ and $K_{1B}$ due to the chiral symmetry breaking effect associated with the significant difference between the $u$ and $s$ quark masses \cite{Suzuki:1993yc, Volkov:2019awd}. This mixing has the form
	\begin{eqnarray}
	\label{K1AK1B}
	K_1(1270) & = & K_{1A}\sin{\alpha} + K_{1B}\cos{\alpha}, \nonumber\\
	K_1(1400) & = & K_{1A}\cos{\alpha} - K_{1B}\sin{\alpha}.
	\end{eqnarray} 

The meson $K_{1}$ appearing in the Lagrangian (\ref{Lagrangian}) corresponds to the state $K_{1A}$. For the state $K_{1B}$ which is not described by the NJL model we apply the following vertex of its interaction with quarks:
\begin{eqnarray}
    L = \frac{g_{B}}{2} \sum_{j=0,\pm} K_{1B}^{\mu j} \left(\bar{q} \lambda_{j}^{K} \gamma^{5} \stackrel{\leftrightarrow}{\partial_{\mu}} q\right),
\end{eqnarray}
where the coupling constant is defined in the following way \cite{Volkov:2019awd}:
\begin{eqnarray}
    g_{B} = \left(I_{10} + I_{01}\right)^{-1/2}.
\end{eqnarray}

As a result, after taking into account the contributions of the axial vector states $K_{1A}$ and $K_{1B}$, for the total decay amplitude of $\tau \to K_1(1270) \pi \nu_\tau$ we obtain:
    \begin{eqnarray}
    \label{amplitude_k1pi}
    \mathcal{M}(\tau^- \to K_1(1270)^- \pi^0 \nu_\tau) = -G_F V_{us} g_\pi L_\mu {\left[i \mathcal{M}_{K_{1A}} + \mathcal{M}_{K_{1B}} \right]}^{\mu\nu} \epsilon_\nu(p_{K_1}). 
    \end{eqnarray} 
    
The separate contributions from the axial-vector states have the form
    \begin{eqnarray}
    \label{amplitude_k1pi1}
    \mathcal{M}_{K_{1A}}^{\mu\nu} & = & 2m_s \sin{\alpha} \left[ I_{11}^{K_1}g^{\mu\nu} + I_{11}^{K_1 K^*} \frac{C_{K^*}}{g_{K^*}} \frac{g^{\mu\nu}s f(s) - p^\mu p^\nu f(M^2_{K^*})}{M^2_{K^*} - s - i \sqrt{s}\Gamma_{K^*}} \right. \nonumber\\
    && \left. + I_{11}^{K_1 K^{*'}} \frac{C_{K^{*'}}}{g_{K^*}} \frac{g^{\mu\nu}s f(s) - p^\mu p^\nu f(M^2_{K^{*'}})}{M^2_{K^{*'}} - s - i \sqrt{s}\Gamma_{K^{*'}}}\right],
    \end{eqnarray} 
    \begin{eqnarray}
    \label{amplitude_k1pi2}
    \mathcal{M}_{K_{1B}}^{\mu\nu} = g_{B} g_{\pi} (I_{10}-m^2_s I_{11}) \cos{\alpha} \left[g^{\mu\nu} + \frac{g^{\mu\nu}s f(s) - p^\mu p^\nu f(M^2_{K^*})}{M^2_{K^*} - s - i \sqrt{s}\Gamma_{K^*}} \right],
    \end{eqnarray} 
    where
    \begin{eqnarray}
    f(s)=1- \frac{3}{2}\frac{(m_s-m_u)^2}{s}. 
    \end{eqnarray}  

The constants $C_{K^*}$ and $C_{K^{*'}}$ are defined according to (\ref{C_const}). The factor $R_{K^{*}}$ takes the form
\begin{eqnarray}
    R_{K^*} = \frac{I_{11}^{f}}{\sqrt{I_{11}I_{11}^{ff}}}.
\end{eqnarray}

The $\tau \to K_1(1270) K \nu_\tau$ decay amplitude includes contributions from the contact diagram and a diagram with an intermediate nonstrange $\rho$ meson. For the amplitude of the process under consideration in the extended NJL model, we obtain
    \begin{eqnarray}
    \label{amplitude_k1k}
    \mathcal{M}(\tau^- \to K_1(1270)^- K^0 \nu_\tau) = G_F V_{ud} \sqrt{2} L_\mu {\left[i \mathcal{M}_{K_{1A}} + \mathcal{M}_{K_{1B}} \right]}^{\mu\nu} \epsilon_\nu(p_{K_1}),
    \end{eqnarray}  
    where
    \begin{eqnarray}
    \label{amplitude_k1pi1}
    \mathcal{M}_{K_{1A}}^{\mu\nu} & = & (m_s+m_u) \sin{\alpha} \left[ I_{11}^{K_1K}g^{\mu\nu} + I_{11}^{\rho K_1 K} \frac{C_\rho}{g_\rho} \frac{g^{\mu\nu}s - p^\mu p^\nu}{M^2_\rho - s - i \sqrt{s}\Gamma_\rho} \right. \nonumber\\
    && \left. + I_{11}^{\rho^{'} K_1 K} \frac{C_{\rho^{'}}}{g_\rho} \frac{g^{\mu\nu}s - p^\mu p^\nu}{M^2_{\rho^{'}} - s - i \sqrt{s}\Gamma_{\rho^{'}}}\right].
    \end{eqnarray}
    \begin{eqnarray}
    \label{amplitude_k1pi1}
    \mathcal{M}_{K_{1B}}^{\mu\nu} & = & -g_{B} g_{K}\cos{\alpha} \left[ g^{\mu\nu} +  \frac{g^{\mu\nu}s - p^\mu p^\nu}{M^2_\rho - s - i \sqrt{s}\Gamma_\rho} \right]  \nonumber\\
    && \times \left\{I_{10} - \left[\left(m_{s} - m_{u}\right)^{2} + m_{u}^{2}\right] I_{11} - 2m_{u}^{3}\left(m_{s} - m_{u}\right) I_{21}\right\}.
    \end{eqnarray}  	
	
   The calculated branching fractions for all considered $\tau$ decays into axial-vector and pseudoscalar mesons are given in the Table \ref{Tab}. In this Table, we also compare our numerical estimates with the results of the papers \cite{Calderon:2012zi, Dai:2018zki}.
   
	\begin{table}[h]
		\caption{Numerical estimates of branching fractions.}
		\label{Tab}
		\begin{center}
			\begin{tabular}{c|c|c|c|c}
				Decay mode  & \cite{Calderon:2012zi} & \cite{Dai:2018zki} & NJL \\
				\hline
				$\tau \to a_{1}(1260)^{-} \pi^{0} \nu_{\tau}$ & $6.9 \pm 6.3$  & & 0.14 & $\times 10^{-3}$ \\
				& $6.1 \pm 5.9$  & & & $\times 10^{-3}$ \\
				$\tau \to a_{1}(1260)^{0} \pi^{-} \nu_{\tau}$ & $6.8 \pm 6.1$  & 1.3 & 0.13 & $\times 10^{-3}$\\
				& $5.9 \pm 5.7$  & & & $\times 10^{-3}$\\
				$\tau \to K_{1}(1270)^{-} \pi^{0} \nu_{\tau}$ & $0.8 \pm 0.2$ & & 3.59 & $\times 10^{-6}$\\
				$\tau \to K_{1}(1270)^{0} \pi^{-} \nu_{\tau}$ & $1.4 \pm 0.5$ & 21 & 6.84 & $\times 10^{-6}$\\
				$\tau \to K_{1}(1400)^{-} \pi^{0} \nu_{\tau}$ & $1.1 \pm 0.1$ & & 0.27 & $\times 10^{-6}$\\
				$\tau \to K_{1}(1400)^{0} \pi^{-} \nu_{\tau}$ & $2.1 \pm 0.2$ & 4.1 & 0.49 & $\times 10^{-6}$\\
				$\tau \to K_{1}(1270)^{-} K^{0} \nu_{\tau}$ & $2.8 \pm 1.9$ & & 4.25 & $\times 10^{-9}$\\
				$\tau \to K_{1}(1270)^{0} K^{-} \nu_{\tau}$ & $13 \pm 8.9$ & & 6.79 & $\times 10^{-9}$\\
			\end{tabular}
		\end{center}
	\end{table}

	\begin{figure}[h]
		\center{\includegraphics[scale = 0.8]{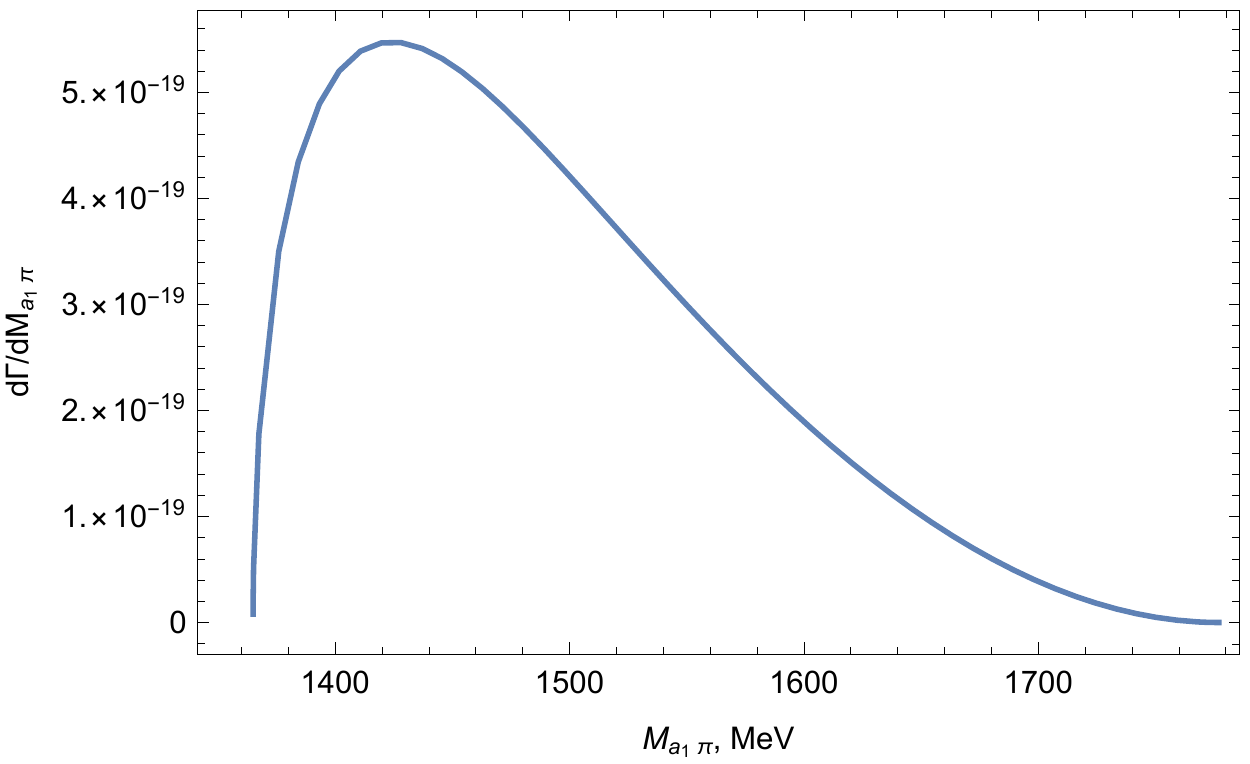}}
		\caption{The invariant mass distribution for $\tau^- \to a^-_1 \pi^0 \nu_\tau$ decay.}
		\label{a1pi}
	\end{figure}

	\begin{figure}[h]
		\center{\includegraphics[scale = 0.8]{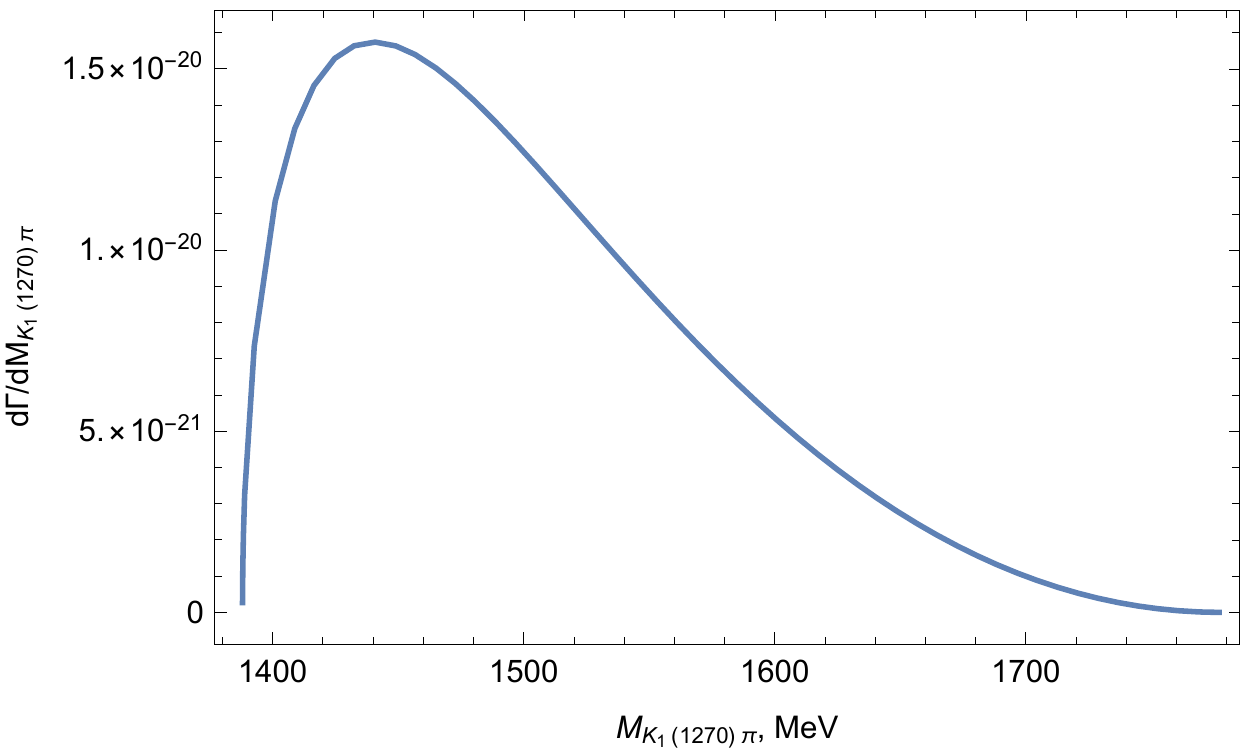}}
		\caption{The invariant mass distribution for $\tau^- \to K^-_1(1270) \pi^0 \nu_\tau$ decay.}
		\label{k1pi}
	\end{figure}
	
		\begin{figure}[h]
		\center{\includegraphics[scale = 0.8]{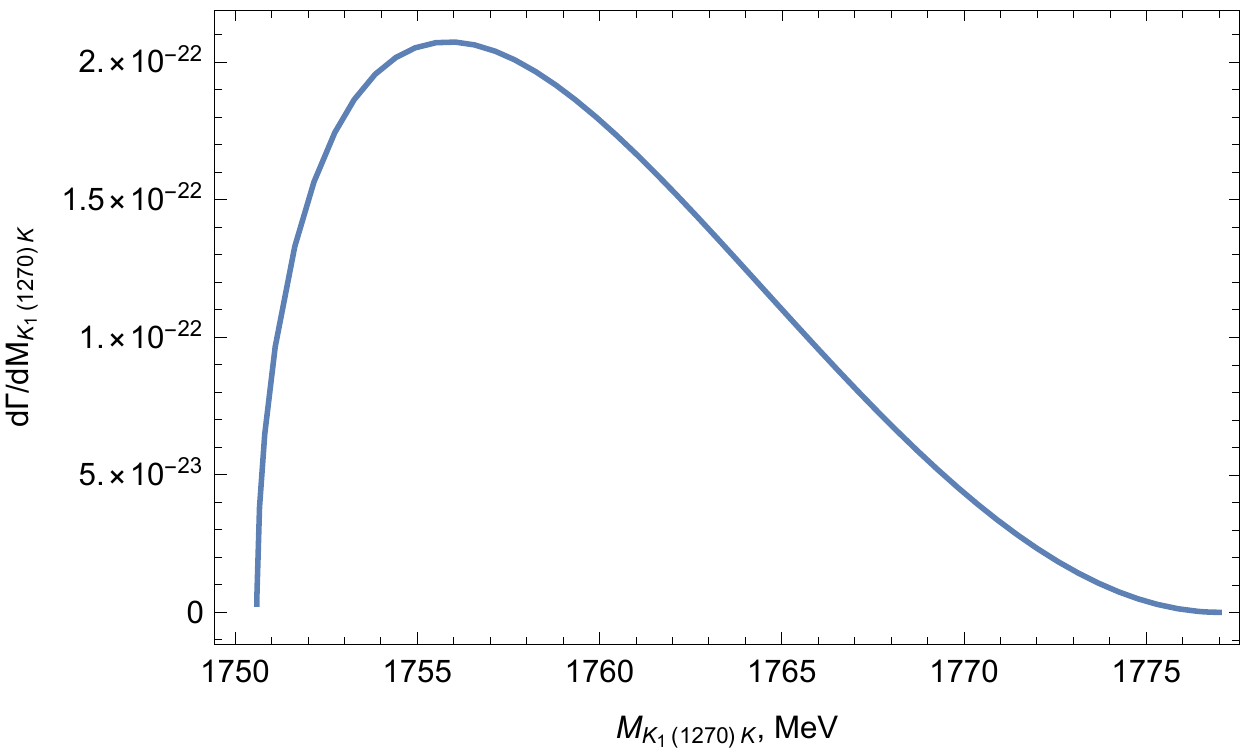}}
		\caption{The invariant mass distribution for $\tau^- \to K^-_1 K^0 \nu_\tau$ decay.}
		\label{k1k}
	\end{figure}

\section{Discussion and Conclusion}
As we can see our results are noticeably different from the results obtained in \cite{Calderon:2012zi, Dai:2018zki} (see Table \ref{Tab}). Our predictions for branching fractions are much smaller than those obtained in paper \cite{Dai:2018zki} and some predictions from the paper \cite{Calderon:2012zi}. 

Our calculations show that the branching fractions of $\tau \to a_1 \pi \nu_\tau$ decay turns out to be sensitive to the option of the axial vector $a_1$ meson mass.
When calculating the decay widths, we use the values of mesons masses and widths in accordance with PDG \cite{ParticleDataGroup:2020ssz}. In particular, the mass of the nonstrange axial vector meson is defined as $M_{a_1}= 1230 \pm 40$ MeV. At the same time, in the NJL model the mass formula for the $a_1$ meson $M^2_{a_1}=M^2_\rho + 6m^2_u$ gives the value $M_{a_1} \approx 1020$ MeV. It is interesting to note that a close value of this mass $M_{a_1}=998(49)$ MeV was recently obtained in the study of the axial-vector form factor in the radiative pion decay \cite{Mateu:2007tr}. In both these cases, the $a_{1}$ meson mass turns out to be close to 1 GeV. Using this mass value the branching fractions of $\tau \to a^-_1 \pi^0 \nu_\tau$ decay increases to $Br(\tau \to a^-_1 \pi^0 \nu_\tau) = 1.45 \times 10^{-3}$ in our model. 
It should be noted that in the paper \cite{Dumm:2009va} the authors obtained an estimate of the $a_1$ meson mass $M_{a_1}=1120$ MeV from the analysis of the three-pion tau-lepton decay experiment of the ALEPH collaboration. These are purely qualitative estimates that do not take into account the changes in the constants associated with the option of the $a_1$ meson mass.

Our result for the decay $\tau \to a_{1}(1260)^{-} \pi^{0} \nu_{\tau}$ is much smaller than calculated in \cite{Calderon:2012zi}, at the same time, $\tau \to K_{1}(1270)^{0} \pi^{-} \nu_{\tau}$ and $\tau \to K_{1}( 1270)^{-} K^{0} \nu_{\tau}$ processes branching fractions are in the same order. This may be due to the fact that in our model we took into account the lower-order terms of the expansion in terms of external momentum. This approach differs from the methods, which used in VDM \cite{Calderon:2012zi}. Indeed, the NJL model takes into account only logarithmically divergent terms in quark loops, where there are minimal degrees of dependence on the external momentum. This approach corresponds to the requirement of maximum preservation of the approximate chiral symmetry in the model, which makes it possible to use the minimum number of arbitrary parameters. In this matter, our approach notably differs from the methods used by other authors. Besides, the branching fractions of several decays obtained in \cite{Calderon:2012zi} have too big uncertainties. The uncertainties of calculations carried out in the $U(3) \times U(3)$ chiral NJL model can be estimated at $15\%$ \cite{Volkov:2022jfr}.

The invariant mass distribution for the processes $\tau^- \to a^-_1 \pi^0 \nu_\tau$, $\tau^- \to K^-_1(1270) \pi^0 \nu_\tau$ and $\tau^- \to K^-_1 K^0 \nu_\tau$ obtained in this work are presented in Fig. \ref{a1pi}, \ref{k1pi} and \ref{k1k}. Our peak for the decay $\tau^- \to a^-_1 \pi^0 \nu_\tau$ is shifted to the left compared with the peak from \cite{Dai:2018zki}. Besides, in \cite{Dai:2018zki}, the invariant mass distribution begins before the threshold and grows much smoother than in our case. For the other processes the behavior is the same.

Our method led to quite satisfactory results in describing a significant number of $\tau$ decays with the production of meson pair containing vector and pseudoscalar ones. In particular, this applies to decays with the production of such meson pairs as $\rho\pi$, $K^* \pi$ and $K^* K$ \cite{Volkov:2022jfr}. In these decays, the main role is played by the same triangular vertex containing axial-vector, vector, and pseudoscalar mesons, as in the case of the considered here decays.
 
\subsection*{Acknowledgments}
The authors are grateful to prof. A. B. Arbuzov for fruitful  discussions. This research has been funded by the Science Committee of the Ministry of Education and Science of the Republic of Kazakhstan (Grant No. AP09057862).

\end{document}